\documentclass[a4paper]{jpconf}
\usepackage{graphicx}
\usepackage{amsmath,amssymb,physics}

\newcommand{\Sch}{Schr\"{o}dinger}
\newcommand{\Heff}{H_\textrm{eff}}
\newcommand{\sgn}{\operatorname{sgn}}

\begin{document}
\title{What is the resonant state in open quantum systems?}

\author{Naomichi Hatano}

\address{Institute of Industrial Science, The University of Tokyo, 5-1-5 Kashiwanoha, Kashiwa, Chiba 277-8574, Japan}

\ead{hatano@iis.u-tokyo.ac.jp}

\begin{abstract}
The article reviews the theory of open quantum system from a perspective of the non-Hermiticity that emerges from the environment with an infinite number of degrees of freedom.
The non-Hermiticity produces resonant states with complex eigenvalues, resulting in peak structures in scattering amplitudes and transport coefficients.
After introducing the definition of resonant states with complex eigenvalues, 
we answer typical questions regarding the non-Hermiticity of open quantum systems.
What is the physical meaning of the complex eigenmomenta and eigenenergies?
How and why do the resonant states break the time-reversal symmetry that the system observes?
Can we make the probabilistic interpretation of the resonant states with diverging wave functions?
What is the physical meaning of the divergence of the wave functions?
We also present an alternative way of finding resonant states, namely the Feshbach formalism, in which
we eliminate the infinite number of the environmental degrees of freedom.
In this formalism, we attribute the non-Hermiticity to the introduction of the retarded and advanced Green's functions.
\end{abstract}

\section{Introduction: What is the open quantum system?}
\label{sec1}

Open quantum systems are called open in the sense that they exchange particle and energy fluxes with the outside world. 
The open system is often referred to as the system when the outside world is called the environment.

As a typical example, consider the nuclear scattering problem of radiating neutrons to a nuclide.
The nuclide may absorb a neutron, becoming a radioactive isotope, and after its lifetime, it may emit a neutron, becoming back to the original nuclide.
We can regard this as an open quantum system, in which the nuclide is the system and the free space with flying neutrons is the environment.

This is a typical setup of the problem of open quantum systems;
the system of interest is often a complex but finite system that is embedded in the environment which is a simple and flat space with an infinite number of degrees of freedom.
Another example found in condensed-matter physics is a quantum dot attached to quantum leads set in semiconductor hetero-junctions.
Indeed, all quantum systems used in experiments must be open because we cannot help making a macroscopic probe interact with the system.

In conventional physics, however, theories often analyze closed systems because it is much easier to treat.
As such, experimentalists who wish to reproduce the theoretical setup often must pay extra attention to keeping the system as closed and undisturbed as possible.
Physics of open systems lets us take an alternative way;
we try to explore novel phenomena that emerge due to the openness of the system.
Experimentalists then can take advantage of the fact that they attach experimental apparatus to the system of interest.

Characteristic phenomena that emerge in open quantum systems include the non-Hermiticity and resonance.
The effective Hamiltonian of open systems can be non-Hermitian;
indeed, the probability amplitude and the energy are not conserved due to the interaction with the environment, and hence the Hamiltonian should be non-Hermitian.
The Feshbach formalism~\cite{Feshbach58review,Feshbach58,Feshbach62}, which we will review later in Section~\ref{sec3}, attributes the appearance of the non-Hermiticity to the elimination of the infinite degrees of freedom of the environment.

Since the Hamiltonian is non-Hermitian, it can harbor eigenstates with complex eigenvalues.
In fact, such eigenstates are none other than the resonant states~\cite{Hatano08,Hatano10,Hatano11,Sasada11,Klaiman11,Hatano13,Hatano14}.
The imaginary part of the complex eigenvalue corresponds to the inverse lifetime of the resonance.
In this framework of understanding open quantum systems, the radioactive isotope is indeed a resonant state of the nuclide with a complex eigenvalue whose imaginary part is equal to half the inverse lifetime of the isotope.

\begin{figure}
\centering
\includegraphics[width=0.82\textwidth]{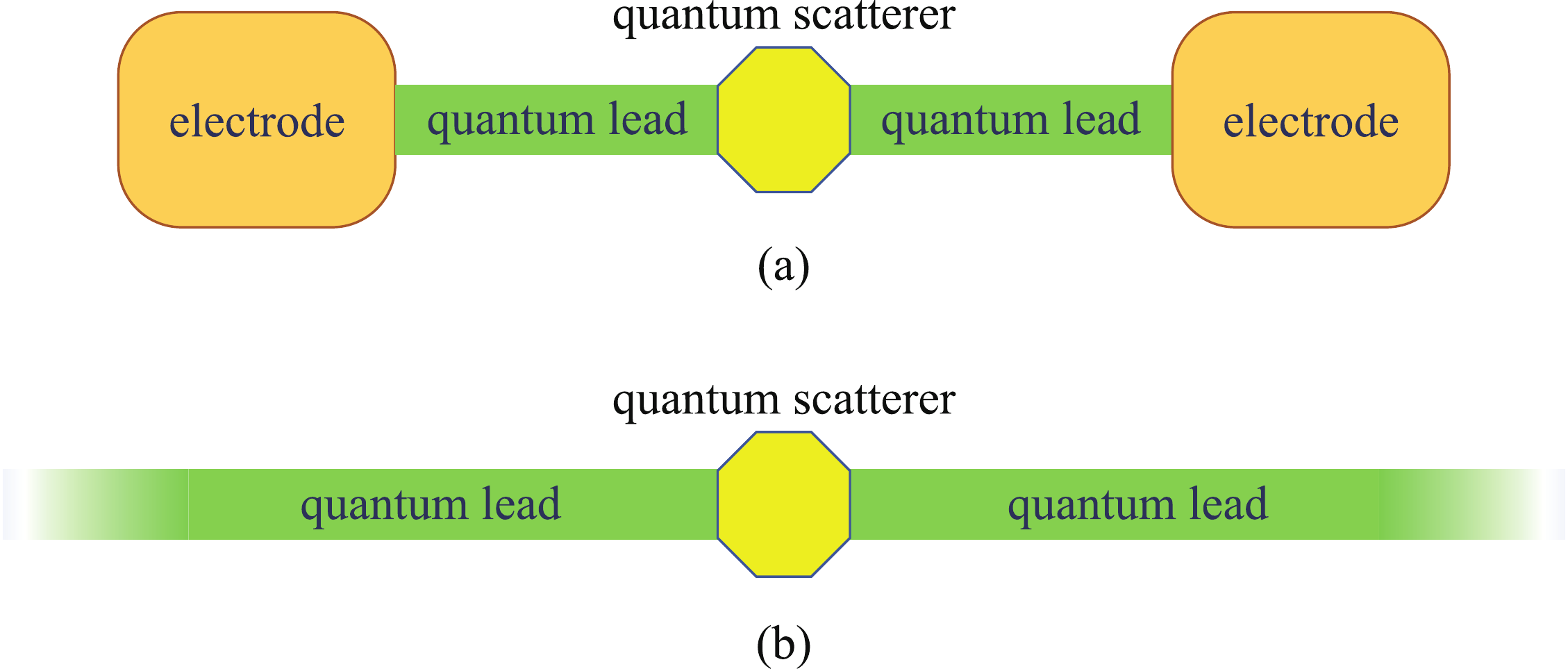}
\caption{(a) Two electrodes are attached to a quantum scatterer by way of quantum leads without impurities.
(b) The potential scattering problem of the quantum scatterer.
}
\label{fig1}
\end{figure}

Another incident in which the resonant state manifests itself is the peak structure of transport coefficients.
Suppose that one tries to measure the electric conductance of a quantum scatterer.
One would attach two electrodes to the scatterer with quantum leads without impurities; see Fig.~\ref{fig1}(a).
Electrons that one electrode emits are transported ballistically through the quantum lead, and are scattered quantum mechanically at the scatterer.
While the reflected electrons go back to the original electrode, the transmitted electrons go to the other electrode, contributing to the electric conduction.
The Landauer formula~\cite{Landauer57,Datta95} dictates that the transmission coefficient of the quantum scattering problem gives the electric conductance $G$ in the form
\begin{align}\label{eq10}
G=\frac{2e^2}{h}T(E_\textrm{F}),
\end{align}
where $h$ denotes the Planck constant and $T(E_\textrm{F})$ is the transmission coefficient at the Fermi energy $E_\textrm{F}$ of the quantum scatterer.

One might wonder why the electric conductivity of the finite section that are sandwiched by the two electrodes as in Fig.~\ref{fig1}(a) is given by solving the scattering problem that are set in an infinite space as in Fig~\ref{fig1}(b).
It is because each electrode has by definition an infinite number of degrees of freedom, and hence electrons that are absorbed into the electrode will never come out of it coherently.
This situation is the same as in the quantum scattering problem, in which the reflected and transmitted waves leave to the negative and positive infinites, and will never come back to the scattering area.
For this reason, the Landauer formula~\eqref{eq10} lets us measure the transmission coefficient in solid-state setups, which may exhibit a resonant peak whose width is proportional to the imaginary part of the complex eigenvalue of the resonant state.

We will come back to this point later in Subsection~\ref{subsec2.6};
we will attribute the fact that the resonant wave function diverges in the limit $x\to\infty$ in the setup of Fig.~\ref{fig1}(b) to the fact that the electrodes contain a macroscopic number of electrons in the setup of Fig.~\ref{fig1}(a).

\section{What is the resonant state?}
\label{sec2}

\subsection{Dynamical and static pictures of resonance phenomena}
\label{subsec2.1}

Let us first note two different ways of understanding resonance.
One is dynamic and the other is static.

In the dynamic picture of resonance phenomena, one would consider the time evolution of an initial state in which a wave packet comes into a scattering potential.
The wave packet would be then captured by a scattering potential,  stay there during the resonance lifetime, and  tunnel out of the potential.
This picture is particularly effective when the system is weakly coupled to the environment.
The system without coupling would have bound states.
The system with weak coupling then would capture the wave packet well if the energy of the wave packet `resonates' with a bound state.
This would result in a peak in the scattering amplitude.
This picture of the resonance, however, does not work well once the system-environment coupling becomes strong;
the resonance peak would shift from the bound-state energy considerably.

We here instead focus on the static picture of resonance;
we describe the resonance in terms of resonance eigenvalues.
In order to emphasize the difference,
let us give an example of coupled pendulums from classical mechanics.
The equation of motion takes the form
\begin{align}\label{eq5}
\ddot{x}_1&=-\omega^2 x_1 - \alpha (x_1-x_2),
\\\label{eq6}
\ddot{x}_2&=-\omega^2 x_2 -\alpha (x_2-x_1),
\end{align}
where $x_1$ and $x_2$ denote the amplitudes of the oscillation of the respective pendulums, $\omega$ denotes their common eigenfrequency, and $\alpha$ denotes the coupling between them.
In the dynamic picture, we understand the coupled pendulums in the following way.
When we start the time evolution from the initial state in which only one of the pendulum, say $x_1$,  oscillates, the oscillation diminishes gradually while the oscillation of the other pendulum, $x_2$, is enhanced.
After the oscillation of the first pendulum stops and that of the second one reaches its maximum, the time evolution starts going back to the initial state.
This is a familiar description of the double pendulum.
Indeed, it is the dynamical picture of this resonance phenomenon.
In fact, the initial state is \textit{not} an eigenstate of the dynamics, and hence is unstable.

The static  picture instead offers the following description.
We find the eigenstates by casting Eqs.~\eqref{eq5}-\eqref{eq6} into the matrix form
\begin{align}\label{eq7}
\dv[2]{t}\mqty(x_1 \\ x_2 )
=-\mqty(
\omega^2 +\alpha & -\alpha \\
-\alpha & \omega^2+\alpha
)\mqty(x_1 \\ x_2 ),
\end{align}
and by diagonalizing the matrix on the right-hand side.
The eigenstates are given as follows:
\begin{align}
\mqty(1 \\ 1) &\quad\mbox{for the eigenvalue $\omega^2$},\\
\mqty(1 \\ -1)&\quad\mbox{for the eigenvalue $\omega^2+2\alpha$},
\end{align}
The first eigenstate indicates the state in which the two pendulums oscillate with the same phase as in $x_1=x_2$ while the second eigenstate indicates the one in which they oscillate with the phase difference $\pi$ as in $x_1=-x_2$.
Their eigenvalues tell us that the first mode oscillates at the original frequency while the second one does at a shifted frequency.
\textit{This} is the static picture of the resonance of the double pendulum.
The dynamic picture is recovered by superimposing the two eigenstates accompanied by time-evolving phases.

In the following, we describe quantum resonances in terms of the latter static picture.
We find resonant eigenstates and eigenvalues of the standard potential scattering problem of the time-independent \Sch equation in one dimension:
\begin{align}\label{eq80}
H\psi(x)=E\psi(x)
\end{align}
with the Hamiltonian
\begin{align}\label{eq81}
H=-\frac{\hbar^2}{2m}\dv[2]{x}+V(x)
\end{align}
where $V(x)$ is a real function that exists only in a finite region $-\ell\leq x \leq \ell$.

In the following subsections, we first present in Subsect.~\ref{subsec2.2} the definition of resonance as an eigenstate of the \Sch equation under a special boundary condition.
The well-known eigenstates of the \Sch equation are the bound states and the scattering states.
The bound states are normalizable, while the scattering states are box-normalizable and conserve the probability flux.
In addition to these eigenstates, we here define unnormalizable eigenstates of the \Sch equation, called the resonant states, the anti-resonant states and the anti-bound states.
The first two in particular are the solutions with broken time-reversal symmetry, which we discuss in Subsect.~\ref{subsec2.3}, and with complex eigenvalues, which we discuss in Subsect.~\ref{subsec2.4}.
Finally, we consider the unnormalizability of the resonant and anti-resonant states.
We will show in Subsect.~\ref{subsec2.5} that the unnormalizability is in fact a necessary condition for the probability conservation;
the wave functions of these states diverge in space but decay in time, and thereby conserves the probability.
We also consider in Subsect.~\ref{subsec2.6} the unnormalizability from the viewpoint of the Landauer formula in Eq.~\eqref{eq10} and in Fig.~\ref{fig1}.

\subsection{Defining the resonance as an eigenstate of the \Sch equation}
\label{subsec2.2}

A textbook definition of the resonance is the pole of the scattering amplitude.
Let us set the following boundary conditions on the scattering wave function $\psi(x)$:
\begin{align}\label{eq20}
\psi(x)=\begin{cases}
Ae^{ikx}+Be^{-ikx} & \quad\mbox{for $x\leq-\ell$},\\
Ce^{ikx} & \quad\mbox{for $x\geq \ell$},
\end{cases}
\end{align}
where the positive wave number $k$ is related to the incident energy $E$ as in $E=\hbar^2 k^2/(2m)$, while the amplitudes $A$, $B$ and $C$ are generally complex numbers that are fixed as functions of $E$ after solving the connection conditions of the wave functions.
The reflection and transmission probabilities are then given by the energy-dependent functions $R(E)=\abs{B/A}^2$ and $T(E)=\abs{C/A}^2$, respectively.
They must be in the range of $[0,1]$ for $E>0$, but can diverge at multiple points on the complex $E$ plane, which most textbooks define as the resonant states.

The poles of $R$ and $T$ (in the complex $E$ plain) are actually identical to the zeros of the incident-wave amplitude $A(E)$.
This observation lets us define the boundary conditions for the resonant eigenstates by removing the incident wave from Eq.~\eqref{eq20}:
\begin{align}\label{eq30}
\psi(x)=\begin{cases}
Be^{-ikx} & \quad\mbox{for $x\leq-\ell$},\\
Ce^{ikx} & \quad\mbox{for $x\geq \ell$},
\end{cases}
\end{align}
which is indeed called the Siegert boundary condition.
We have outgoing waves only if $\Re k>0$ and incoming waves only if $\Re k<0$.

\begin{figure}
\centering
\includegraphics[width=0.8\textwidth]{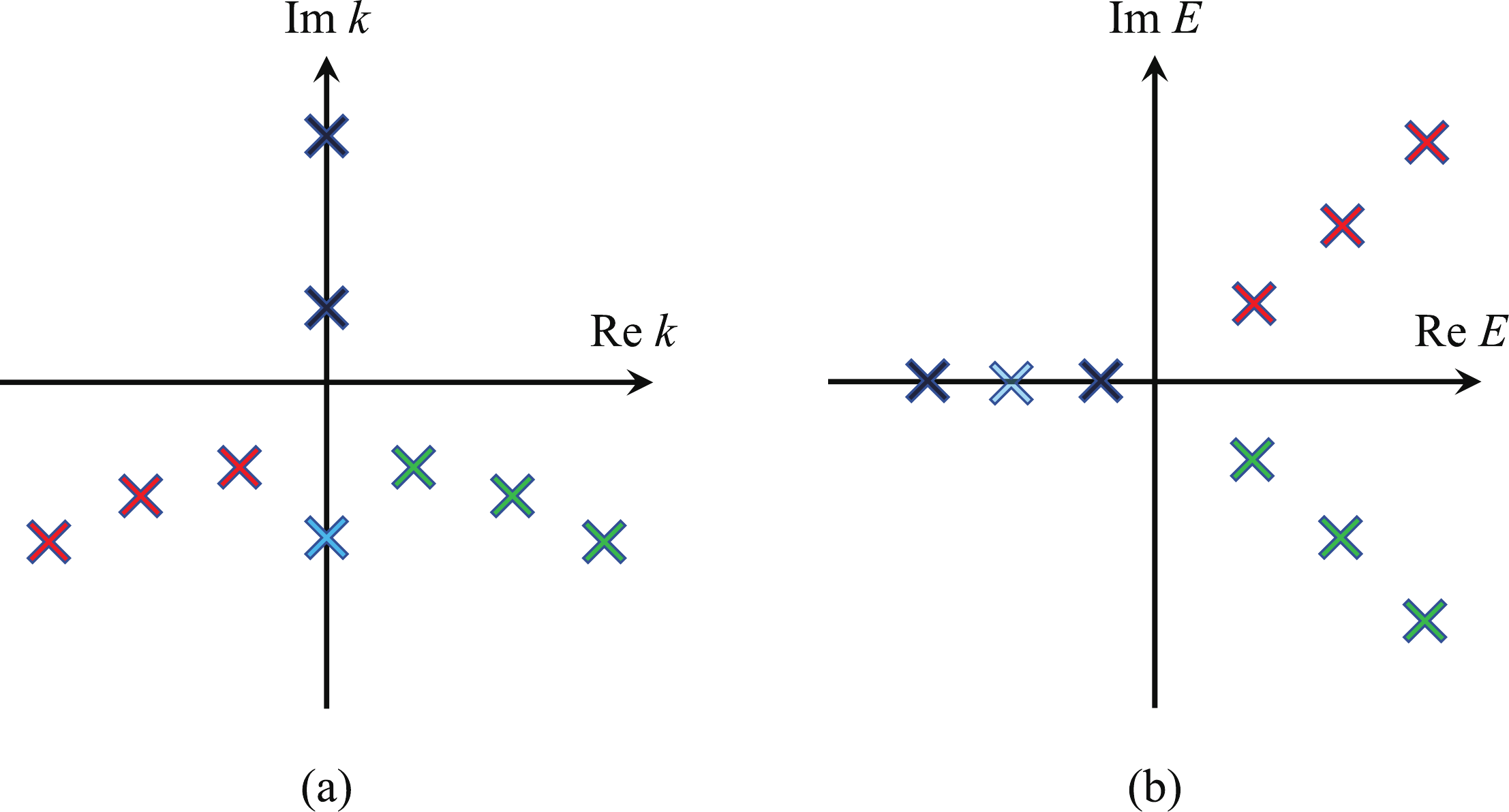}
\caption{The point spectra for the scattering problem under the boundary (a) in the complex $k$ plain and (b) in the complex $E$ plain.
In the complex $k$ plain (a), the bound states are located on the positive imaginary axis, the anti-bound states are on the negative imaginary axis, the resonant states are in the fourth quadrant, and the anti-resonant states are in the third quadrant.
In the complex $E$ plain (b), the bound states are on the negative real axis in the first Riemann sheet, the anti-bound states are on the negative real axis in the second Riemann sheet, the resonant states are in the lower half of the second Riemann sheet, and the anti-resonant states are in the upper half of the second Riemann sheet.}
\label{fig2}
\end{figure}

In fact, the Siegert boundary condition~\eqref{eq30} locates all point spectra, or all the discrete eigenvalues.
Figure~\ref{fig2} shows a schematic picture of a typical distribution of discrete eigenvalues on the complex $k$ plain and the complex $E$ plain.
The eigenstates located on the positive imaginary axis of the complex $k$ plain are the bound states.
Indeed, inserting $k=i\kappa$ with $\kappa>0$ into the Siegert boundary condition~\eqref{eq30} makes it the exponentially decaying wave function of a bound state.
The corresponding eigenvalue $E=-\hbar^2\kappa^2/(2m)$ is a real negative number, which is also consistent with our standard understanding of the bound state.

The eigenstates located in the fourth quadrant of the complex $k$ plain are called the resonant states, while those in the third quadrant are called the anti-resonant states, presumably for historical reasons.
The energy eigenvalues of the resonant states are located in the lower half of the second Riemann sheet of the complex $E$ plain, while those of the anti-resonant states are in its upper half.
We will explore the physical significance of these eigenvalues in the following subsections.

In the meantime, there are also eigenvalues on the negative imaginary axis of the complex $k$ plain.
These states are often called the anti-bound states or the virtual bound states.
The corresponding energy eigenvalues are located on the negative real axis of the second Riemann sheet of the complex $E$ plain.
We will not discuss them much in the present article because they scarcely contribute to physical observables.

Let us mention here the fact that the wave functions of the resonant, anti-resonant and anti-bound states diverge in space.
For these states, inserting $\Im k=i\kappa$ with $\kappa<0$ into the Siegert boundary condition~\eqref{eq30} reveal thethey exponential divergence of the form $\abs{\psi(x)}\sim \exp(\abs{\kappa}\abs{x})$.
This is presumably the reason why these states are sometime called unphysical.
We hereafter try to convince the readers that they are totally physical objects.

A remark on the point spectra and the continuum spectrum is in order here.
Let us assume the simplest scattering problem of a square-well potential:
\begin{align}
V(x)=
\begin{cases}
-V_0 & \quad\mbox{for $\abs{x}\leq \ell$},\\
0 & \quad\mbox{for $\abs{x}> \ell$},
\end{cases}
\end{align}
where $V_0>0$.
We then set the wave function $Fe^{ik'x}+Ge^{-ik'x}$ with ${k'}^2=k^2+2mV_0/\hbar^2$ for the region $[-\ell,\ell]$.
In finding the scattering wave function~\eqref{eq20}, we demand two connection conditions for the continuity of the wave function and its derivative at each point of $x=\pm\ell$, which amounts to four conditions.
Meanwhile we have five unknown variables, namely, the four ratios $A/B$, $C/B$, $F/B$, $G/B$ and the incident energy $E$ (which determines $k$ and $k'$).
We therefore cannot fix all unknown variables.
This is why we have continuous spectrum of $E$ for the scattering wave function~\eqref{eq20}.
In finding the wave function~\eqref{eq30} under the Siegert boundary condition, on the other hand, one unknown variable $A$ is missing.
We demand four conditions to fix four unknown variables including $E$.
This is why we have point spectra of $E$ for the wave function~\eqref{eq30} as exemplified in Fig.~\ref{fig2}.

Let us now analyze the resonant and anti-resonant states by adding the time-dependent factor to the wave function in Eq.~\eqref{eq30}:
\begin{align}\label{eq100}
\Psi(x,t)=e^{-iEt/\hbar}\psi(x)=e^{-iEt/\hbar}\times\begin{cases}
Be^{-ikx} & \quad\mbox{for $x\leq-\ell$},\\
Ce^{ikx} & \quad\mbox{for $x\geq \ell$}
\end{cases}
\end{align}
which is the solution of the time-dependent  \Sch equation
\begin{align}\label{eq110}
i\hbar\pdv{t}\Psi(x,t)=H\Psi(x,t);
\end{align}
note, however, that since $E$ is generally complex, the factor $e^{-iEt/\hbar}$ in Eq.~\eqref{eq100} is generally \textit{not} a phase.

Figure~\ref{fig2} tells us that the resonant states satisfy $\Re k>0$ and $\Im E<0$.
Therefore, the wave function~\eqref{eq100} implies the following:
(i) the wave amplitude that has been captured by the potential leaves the scattering region $-\ell\leq x\leq \ell$ out onto the infinity $\pm\infty$; 
(ii) because of this, the wave amplitude decreases in time.
This is the picture that the resonant state indicates as an eigenstate of the \Sch equation.
On the other hand, the anti-resonant states satisfy $\Re k<0$ and $\Im E>0$, for which
Eq.~\eqref{eq100} implies that (iii) the waves gather from the infinity into the scattering region to be captured by the potential, and therefore (iv) the wave amplitude increases in time.

One might wonder in what physical situations the anti-resonant states emerge.
We can answer this question by reminding ourselves of the dynamical picture of the resonance.
The dynamics of the wave amplitude in the scattering region is recovered by superimposing the states of the point spectra.
The resonant states dominate the dynamics in the initial-condition problem, in which
we ask what will happen \textit{after} the wave amplitude is concentrated in the scattering region.
The anti-resonant states, on the other hand, dominate the dynamics in the terminal-condition problem, in which we ask what has happened \textit{before} the wave amplitude is concentrated.
Of course, it feels much more natural to us to ask the initial-condition problem, but it is only because we as human beings have the memory of the past but not the memory of the future.
In other words, the quantum dynamics itself is time-reversal symmetric with respect to the exchange of the past and the future.

Indeed, K.~Murch and his group, in their experiments of weak measurement, accumulated the observation data of quantum trajectories of a two-level system with leaks into the environment, and showed the following in the data analysis.
Averaging the trajectories that have a common initial condition produces the familiar Rabi oscillation with decaying, while averaging the ones that have a common terminal condition produces the Rabi oscillation with increasing amplitude towards the terminal condition.
This is a manifestation of the anti-resonant contributions in the dynamical picture.

\begin{figure}
\centering
\includegraphics[width=0.55\textwidth]{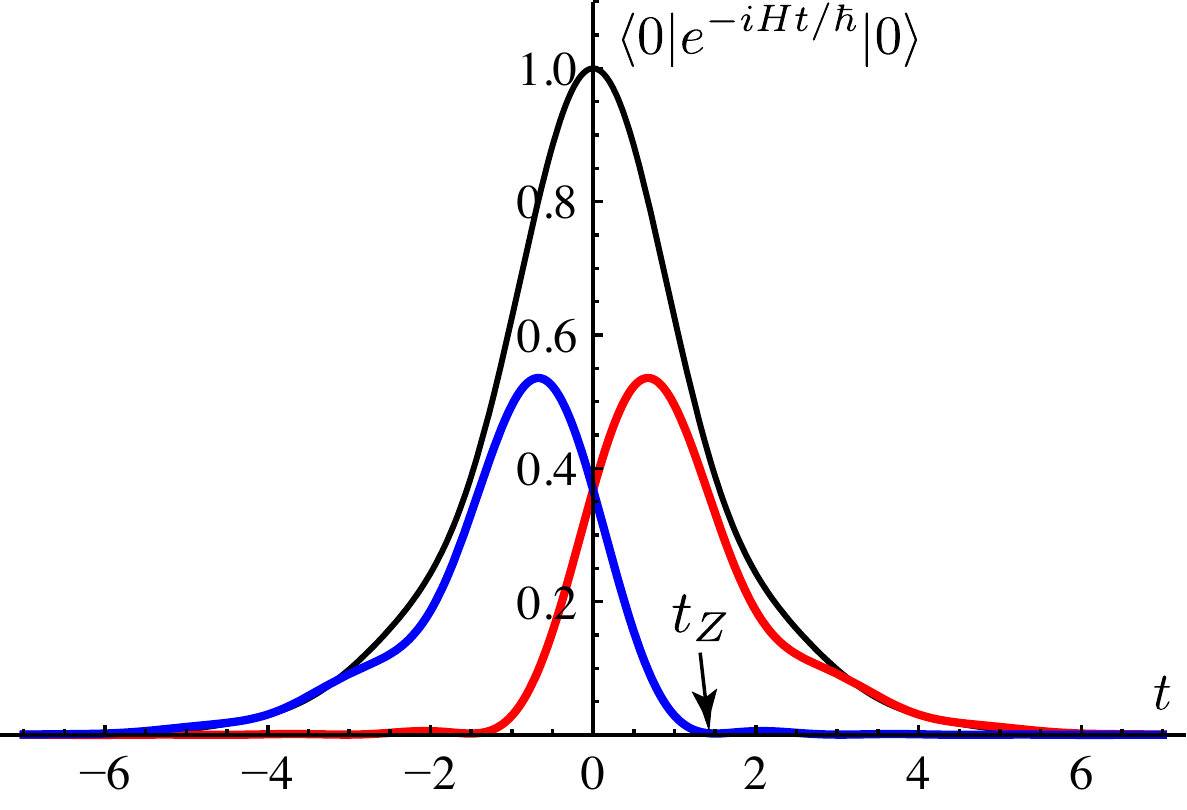}
\caption{Time-dependence of the survival probability of the potential site (denoted by $\ket{0}$ here) of the tight-binding model Hamiltonian $H$.
The black curve with the highest central peak is the total survival probability, while the red curve with the peak on the right is the contribution of the resonant state and the blue curve on the left is the contribution of the anti-resonant state. Taken from Ref.~\cite{Hatano19a}.
The contributions switch from the latter to the former within the quantum Zeno time, which is denoted by $t_Z$ here.}
\label{fig3}
\end{figure}

As another way of explaining the anti-resonant states, we here mention the finding in Ref.~\cite{Hatano14} that all states with the point spectra in Fig.~\ref{fig2} spans the functional space if we restrict the real space $x$ inside the scattering region $-\ell\leq x\leq \ell$.
We can thereby breakdown any dynamics inside the system into the contributions of the states with point spectra including resonant and anti-resonant states.
For example, the survival probability~\cite{Hatano19a} on the potential site at the origin of the tight-binding model of infinite size depends on time as in Fig.~\ref{fig3}.
While the decay of the survival probability for $t>0$ is dominated by the resonant state, the survival probability grows for $t<0$, which is dominated by the anti-resonant state.
The profile is time-reversal symmetric with respect to $t=0$, which reflects the time-reversal symmetry between the resonant and anti-resonant states as we will argue in the next subsection~\ref{subsec2.3}.

\subsection{Broken time-reversal symmetry of the resonant state}
\label{subsec2.3}

Although the dynamics is time-reversal symmetric, we here stress that each of the resonant and anti-resonant states does break the time-reversal symmetry.
Since they appear always as a pair that is symmetric with respect to the imaginary $k$ axis and the real $E$ axis, the system of all solutions indeed keeps the time-reversal symmetry, which the original \Sch equation~\eqref{eq110} accommodates.

One might ask why an eigenstate can break the time-reversal symmetry.
Since the Hamiltonian $H$ in Eq.~\eqref{eq81} is time-reversal symmetric, it commutes with the time-reversal operator, which we denote with $T$ hereafter.
Namely, we have $[H,T]=0$.
One would then claim that all the eigenstates of the Hamiltonian $H$ must be the eigenstates of the time-reversal operator $T$ too, and hence all eigenstates must be time-reversal symmetric.
This deduction is not correct because the time-reversal operator $T$ is \textit{not} a linear operator.

The correct deduction is given as follows.
Suppose that an eigenstate $\ket{\psi_n}$ of the Hamiltonian $H$ has a non-degenerate eigenvalue $E_n$;
that is, $H\ket{\psi_n}=E_n\ket{\psi_n}$.
Applying the time-reversal operator $T$ to both sides of the equation from the left, we have
\begin{align}\label{eq120}
TE_n\ket{\psi_n}=TH\ket{\psi_n}=HT\ket{\psi_n},
\end{align}
where we used the fact $[H,T]=0$ for the second equality.
If $T$ were linear, the left-hand side would be transformed to $E_nT\ket{\psi_n}$, and hence $T\ket{\psi_n}$ would be also an eigenstate of $H$ with the same eigenvalue $E_n$, so that $T\ket{\psi_n}\propto \ket{\psi_n}$.
In truth, $T$ is an anti-linear operator, and hence the left-hand side is transformed to ${E_n}^\ast\, T\ket{\psi_n}$.
We therefore find the following two possibilities:
if $E_n$ is a real eigenvalue, indeed $T\ket{\psi_n}\propto\ket{\psi_n}$, and hence the eigenstate is time-reversal symmetric;
if $E_n$ is a complex eigenvalue, Eq.~\eqref{eq120} tells us that $T\ket{\psi_n}$ is a time-reversal eigenstate of $\ket{\psi_n}$ with a complex-conjugate eigenvalue ${E_n}^\ast$, which implies that we always have a pair of eigenstates that are time-reversal to each other with a complex-conjugate pair of eigenvalues.
The latter possibility is none other than what we observed near the end of the previous subsection~\ref{subsec2.2}.

\subsection{Why does a `Hermitian' Hamiltonian have complex eigenvalues?}
\label{subsec2.4}

One might also wonder why the seemingly Hermitian Hamiltonian $H$ in Eq.~\eqref{eq81} can have complex eigenvalues.
In fact, whether an operator is Hermitian or not depends on what functional space the operator is defined in.
Indeed, the Hamiltonian $H$ is Hermitian in the functional space of bound states and scattering states, and hence these states can only have real eigenvalues, while it is non-Hermitian in the functional space of resonant and anti-resonant states, and hence these states can have complex eigenvalues.

Let us observe in Fig.~\ref{fig2}(a) that $\Im k<0$ for all resonant and anti-resonant states.
Inserting this into the wave function~\eqref{eq30}, we find that it exponentially diverges in $x\to\pm\infty$.
For such functions, the first term of the Hamiltonian $H$ can be non-Hermitian, which we show as follows.
Consider the matrix element 
\begin{align}\label{eq140}
\ev{\dv[2]{x}}{\psi};
\end{align}
showing that the matrix element is real for any functions $\ket{\psi}$ would prove that the operator $\dv*[2]{x}$ is Hermitian.
Note here that we must specify the functional space to find functions $\ket{\psi}$ in, which we will be the focus of the argument below.

For the sake of the argument, we first evaluate
\begin{align}\label{eq150}
\ev{\dv[2]{x}}{\psi}_L:=
\int_{-L}^{L}\psi(x)^\ast \dv[2]{x} \psi(x) dx
\end{align}
for $L>\ell$ and then take the limit $L\to\infty$ whenever it is possible.
We can transform Eq.~\eqref{eq150} by means of partial integration:
\begin{align}\label{eq160}
\ev{\dv[2]{x}}{\psi}_L=
\qty[\psi(x)^\ast \dv{x} \psi(x)]_{x=-L}^L
-\int_{-L}^L \qty(\dv{x}\psi(x))^\ast\qty(\dv{x} \psi(x)) dx.
\end{align}
The second term on the right-hand side is real because the integrand $\abs{\dv*{\psi}{x}}^2$ is real.
The first term, on the other hand, can be complex.

The imaginary part of the first term is actually the real part of
\begin{align}\label{eq170}
\eval{\psi(x)^\ast\hat{p}\psi(x)}_{x=L}-\eval{\psi(x)^\ast\hat{p}\psi(x)}_{x=-L},
\end{align}
where $\hat{p}$ denotes the momentum operator
\begin{align}
\hat{p}:=\frac{\hbar}{i}\dv{x}.
\end{align}
The physical meaning of Eq.~\eqref{eq170} becomes more evident when we evaluate it for potential scattering problems in three dimensions;
Eq.~\eqref{eq150} would then be replaced by the volume integral
\begin{align}
\ev{\laplacian}{\psi}_\Omega:=\int_\Omega \psi(\vec{x})^\ast\laplacian\psi(\vec{x})dV
\end{align}
for a region $\Omega$ that contains the support of the scattering potential, while Eq.~\eqref{eq170} would be replaced by the surface integral
\begin{align}\label{eq200}
\int_{\partial\Omega}\hat{\vec{p}}\cdot\vec{n}\,dS,
\end{align}
where $\vec{n}$ denotes the unit vector that is normal to the surface $\partial\Omega$ of the volume $\Omega$.
Equation~\eqref{eq200} indicates the momentum flux that is going out of the volume $\Omega$, and similarly Eq.~\eqref{eq170} is the momentum flux that is going out of the region $[-L,L]$.

Therefore, Eq.~\eqref{eq170} should vanish in the limit $L\to\infty$ for the bound states as well as for the scattering states~\eqref{eq20}  with the flux conservation $\abs{A}^2=\abs{B}^2+\abs{C}^2$.
Therefore, we can take the limit $L\to\infty$ in Eq.~\eqref{eq150}, and thereby prove that Eq.~\eqref{eq140} is real for any bound states and for any scattering states.
We thus conclude that the Hamiltonian $H$ is Hermitian in the functional spaces of the bound states and the scattering states, and therefore the eigenvalues of these states must be real.

On the other hand, we realize that Eq.~\eqref{eq170} should \textit{not} vanish for the resonant and anti-resonant states, when we remember the fact that the amplitude goes out of the scattering region for the resonant states and comes into the region for the anti-resonant states, as we discussed near the end of Subsection~\ref{subsec2.2}; namely, we have $\abs{B}^2+\abs{C}^2>\abs{A}^2=0$.
In fact, we cannot take the limit $L\to\infty$ in Eq.~\eqref{eq150} because the wave functions for these states spatially diverge in $x\to\pm\infty$, and 
the first term on the right-hand side of Eq.~\eqref{eq160} can have an imaginary part in general.
Our conclusion in this case is that the Hamiltonian $H$ is \textit{non}-Hermitian in the functional spaces of the resonant and anti-resonant states, and hence the eigenvalues for these states are allowed to be complex.

\subsection{Probabilistic interpretation of the resonant state}
\label{subsec2.5}

The answer to the question in the previous subsection~\ref{subsec2.4} might spawn yet another question;
could a spatially diverging wave function of the resonant and anti-resonant states accommodate the probabilistic interpretation of the wave function?
In fact, this may be the reason why many researchers have not paid much attention to the resonant states as eigenstates.
Obviously, the wave functions of the resonant and anti-resonant states are not normalizable.
Yet, we here present a way of probabilistic interpretation of the diverging wave function.
Indeed, the divergence is necessary for the following probabilistic interpretation, being precisely cancelled by the temporal decay.

Let us consider the probability amplitude of a resonant eigenstate $\Psi_n(x,t)$ that is contained in a region $[-L,L]$, where $L>\ell$.
We now remember that for a resonant eigenstate, the probability amplitude leaks out of the scattering region onto $\pm\infty$.
Therefore, in order to analyze the probability conservation for a resonant state, we must chase the leaking probability amplitude by expanding the integration range $[-L,L]$.
We may evaluate the leaking speed as follows;
since the real part of the momentum is $\Re \hbar k_n$, where $(\hbar k_n)^2/(2m)=E_n$, let us guess that  the speed is $\Re\hbar k_n/m$. 
We therefore expand the integration range $[-L,L]$ as in
\begin{align}
L(t)=L_0+\frac{\Re\hbar k_n}{m} t,
\end{align}
where $L_0>\ell$, so that we define
\begin{align}
N_n(t)=\int_{-L(t)}^{L(t)} \abs{\Psi_n(x,t)}^2 dx.
\end{align}

We are now in a position to prove $\dot{N}_n=0$.
We find
\begin{align}\label{eq240}
\dot{N}_n(t)=&
\int_{-L(t)}^{L(t)} \dv{t}\abs{\Psi_n(x,t)}^2 dx
+\dot{L}(t)\qty(\abs{\Psi_n(L(t),t)}^2
+\abs{\Psi_n(-L(t),t)}^2)
\\\label{eq250}
=&
\int_{-L(t)}^{L(t)} \qty(\Psi_n(x,t)^\ast \dv{t}\Psi_n(x,t) +\Psi_n(x,t) \dv{t}\Psi_n(x,t)^\ast )dx
\nonumber\\
&+\frac{\Re \hbar k_n}{m}\qty(\abs{\Psi_n(L(t),t)}^2
+\abs{\Psi_n(-L(t),t)}^2).
\end{align}
We prove that the first term precisely cancels the second term on the right-hand side.

Utilizing the time-dependent \Sch equation~\eqref{eq110}, we can transform the first term as follows:
\begin{align}
&\qty(\mbox{first term on the right-hand side of Eq.~\eqref{eq250}})
\nonumber\\
&=\frac{1}{i\hbar}
\int_{-L(t)}^{L(t)} \qty[\Psi_n(x,t)^\ast H\Psi_n(x,t) - \qty(\Psi_n(x,t)^\ast H\Psi_n(x,t))^\ast ]dx
\\
&=\frac{2}{\hbar}\Im\int_{-L(t)}^{L(t)}\Psi_n(x,t)^\ast H\Psi_n(x,t)dx
\\\label{eq280}
&=-\frac{\hbar}{m}\Im\int_{-L(t)}^{L(t)}\Psi_n(x,t)^\ast \Psi_n''(x,t)dx;
\end{align}
in the last transformation, we used the fact that the potential $V(x)$ is a real function.
After the partial integration of Eq.~\eqref{eq280}, we find
\begin{align}\label{eq290}
\qty(\mbox{Eq.~\eqref{eq280}})
&=-\frac{\hbar}{m}\Im\qty(
\Psi_n(L(t),t)^\ast\Psi_n'(L(t),t)
-\Psi_n(-L(t),t)^\ast\Psi_n'(-L(t),t)
),
\end{align}
where the integral part vanishes because it is purely real.

We further use the fact that the resonant eigenfunction behaves as Eq.~\eqref{eq100} in the region $\abs{x}>\ell$, which transforms Eq.~\eqref{eq290} further as 
\begin{align}
\qty(\mbox{Eq.~\eqref{eq290}})
&=-\frac{\hbar}{m}\Im\qty(
ik_n\abs{\Psi_n(L(t),t)}^2
+ik_n\abs{\Psi_n(-L(t),t)}^2)
\\
&=-\frac{\hbar}{m}\Re k_n\qty(\abs{\Psi_n(L(t),t)}^2
+\abs{\Psi_n(-L(t),t)}^2),
\end{align}
which precisely cancels the second term on the right-hand site of Eq.~\eqref{eq250}.
We thereby conclude that $\dot{N}_n(t)\equiv0$, which proves that the probability amplitude contained in the expanding range $[-L(t),L(t)]$ is conserved, offering us a probability interpretation despite that the eigenfunction itself is not normalizable.

In Eq.~\eqref{eq240}, the first term on the right-hand side describes the temporal exponential decrease of the wave amplitude, while the second term describes the spatial exponential increase.
The fact that these two terms cancel each other implies that the spatial divergence is indeed necessary for the probability conservation of the resonant state;
it is nothing but physical.

\subsection{Physical meaning of the spatially divergent wave function}
\label{subsec2.6}

In the previous subsection~\eqref{subsec2.5}, we confirmed that the spatial divergence of the resonant eigenfunction is indeed necessary for the probability to be conserved.
Nonetheless, one might still wonder what would be the physical meaning of the divergence.

The Landauer formula~\eqref{eq10} gives us a hint. 
Of course, we cannot set up the infinite space of Fig.~\ref{fig1}(b) for the potential scattering problem.
In reality, the space for the scattering waves is terminated by electrodes and other experimental probes.
The Landauer formula implies the equivalence between the two setups.

We then realize that the leaking probability fluxes in the resonant states do not really leave onto $\pm\infty$ in Fig.~\ref{fig1}(b) but end up in the electrodes in Fig.~\ref{fig1}(a).
Indeed, when we try to measure the electric conductance of a quantum scatterer, the electrodes are macroscopic objects, and hence should contain a macroscopic number of electrons, while the quantum scatterer is a microscopic or mesoscopic object, and hence contain only a microscopic number of electrons.
We can thereby understand that the spatial divergence of the resonant eigenfunction physically corresponds to the number of electrons in the electrodes of the order of the Avogadro number.

\section{Feshbach formalism for the complex eigenvalues}
\label{sec3}

\subsection{Nonlinear eigenvalue problem and non-Markovian time evolution}
\label{subsec3.1}

In the previous section~\ref{sec2}, we notice that the Hamiltonian acquires non-Hermiticity because of the Siegert boundary condition~\eqref{eq30}.
In the present section, we present an alternative approach to open quantum systems called the Feshbach formalism, in which we find an explicitly non-Hermitian effective Hamiltonian of the system.

Let $P$ and $Q$ denote the operators that project any states onto the subsystems of the system and the environment, respectively.
If the scattering potential $V(x)$ exists only in the region $[-\ell,\ell]$, then for any wave function $\psi(x)$, 
\begin{align}
P\psi(x)&=\begin{cases}
\psi(x) & \quad\mbox{for $-\ell\leq x\leq \ell$},\\
0& \quad\mbox{for $x<-\ell$ or $x>\ell$},
\end{cases}
\\
Q\psi(x)&=\begin{cases}
0 & \quad\mbox{for $-\ell\leq x\leq \ell$},\\
\psi(x) & \quad\mbox{for $x<-\ell$ or $x>\ell$}.
\end{cases}
\end{align}
We have $P^2=P$, $Q^2=Q$, $PQ=QP=0$, and $P+Q=I$, where $I$ is the identity operator.

What we try to achieve is to find the \Sch equation for the wave function only inside the system, namely $P\ket{\psi}$.
In order to do so, we operate $P$ and $Q$ to the time-independent \Sch equation~\eqref{eq80}:
\begin{align}\label{eq330}
PH(P+Q)\ket{\psi}&=EP\ket{\psi},\\
\label{eq340}
QH(P+Q)\ket{\psi}&=EQ\ket{\psi}.
\end{align}
We can cast these equations into the following forms:
\begin{align}\label{eq350}
PHP\qty(P\ket{\psi}) + PHQ \qty(Q\ket{\psi}) &= E\qty(P\ket{\psi}),\\
\label{eq360}
QHP\qty(P\ket{\psi}) + QHQ \qty(Q\ket{\psi}) &= E\qty(Q\ket{\psi}).
\end{align}
We can represent $Q\ket{\psi}$ in terms of $P\ket{\psi}$ using the second equation~\eqref{eq360}:
\begin{align}
Q\ket{\psi}=\frac{1}{E-QHQ}QHP\qty(P\ket{\psi}).
\end{align}
Inserting this expression into the first equation~\eqref{eq350}, we find
\begin{align}\label{eq380}
\Heff(E) \qty(P\ket{\psi}) = E \qty(P\ket{\psi}),
\end{align}
where
\begin{align}\label{eq390}
\Heff(E):=PHP+PHQ\frac{1}{E-QHQ}QHP.
\end{align}
This is exactly what we tried to achieve.
Note that the eigenvalue problem~\eqref{eq380} is a nonlinear one because the Hamiltonian on its left-hand side is a nonlinear function of $E$.

We can understand the effective Hamiltonian $\Heff$ in Eq.~\eqref{eq390} in the following way.
The first term $PHP$ on the right-hand side of Eq.~\eqref{eq390} is the system Hamiltonian.
The second term, on the other hand, represents the effect of the environmental Hamiltonian $QHQ$.
The wave amplitude in the system, namely $P\ket{\psi}$, is transferred to the environment by the coupling Hamiltonian $QHP$, propagates around the environment under the Green's function $(E-QHQ)^{-1}$ of the environmental Hamiltonian $QHQ$, and then the surviving amplitude is transferred transferred back to the system by the coupling Hamiltonian $PHQ$.

\begin{figure}
\centering
\includegraphics[width=0.45\textwidth]{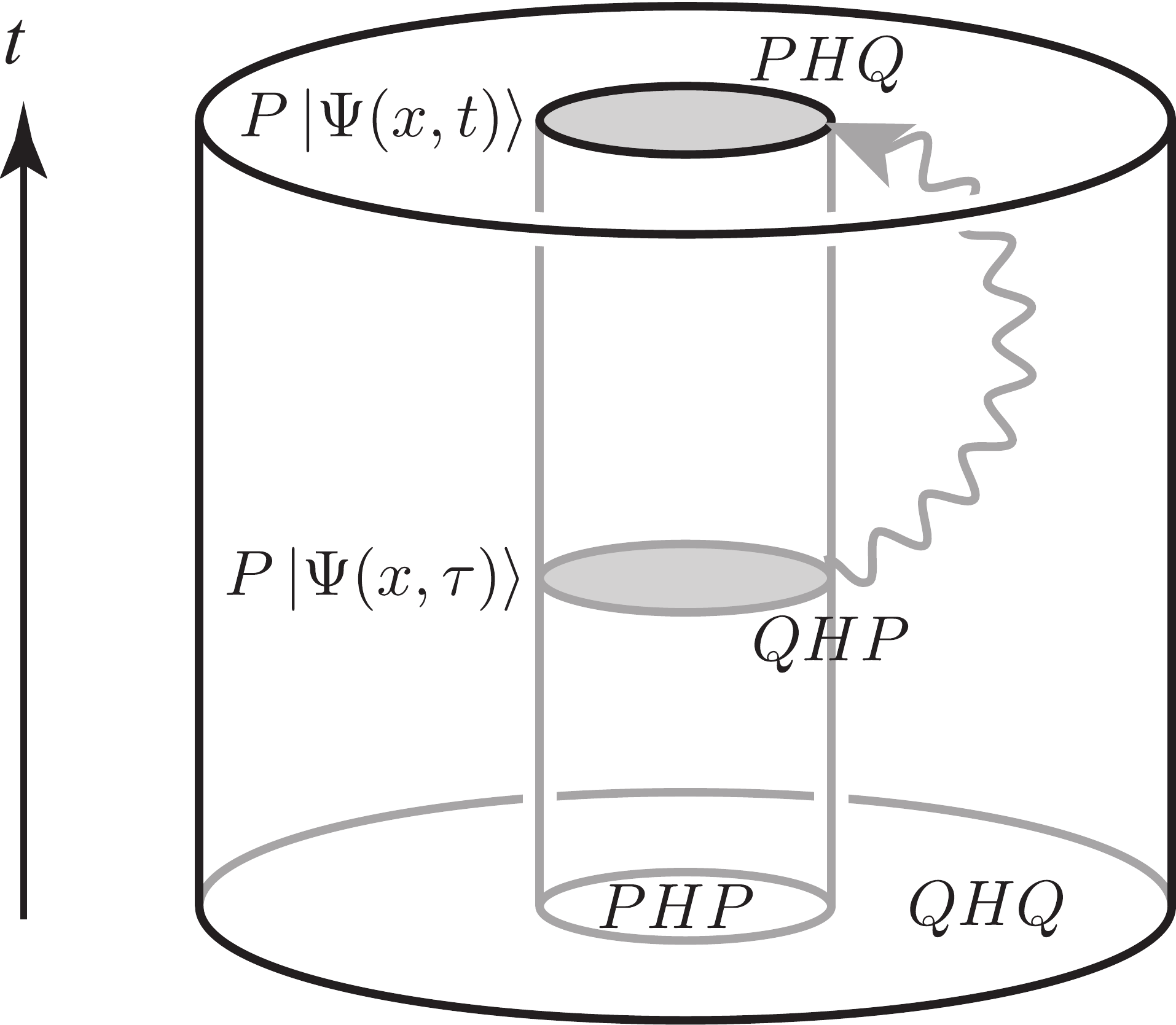}
\caption{A schematic view of the non-Markovian time evolution of the system through the propagation in the environment. The system state $P\ket{\Psi(x,\tau)}$ is transferred to the environment by the coupling Hamiltonian $QHP$, evolves in time during the period $t-\tau$ in the environment, and is transferred back to the system by the coupling Hamiltonian $PHQ$ to be $P\ket{\Psi(x,t)}$, which thus depends on the state in the past indirectly.}
\label{fig4}
\end{figure}

This view becomes more clear when we apply the Feshbach projection formalism to the time-dependent \Sch equation~\eqref{eq110}.
Instead of Eqs.~\eqref{eq350}--\eqref{eq360}, we now have 
\begin{align}\label{eq400}
i\hbar \pdv{t} \qty(P\ket{\Psi(x,t)}) &= PHP\qty(P\ket{\Psi(x,t)})+ PHQ\qty(Q\ket{\Psi(x,t)}),\\
\label{eq410}
i\hbar \pdv{t} \qty(Q\ket{\Psi(x,t)}) &= QHP\qty(P\ket{\Psi(x,t)})+ QHQ\qty(Q\ket{\Psi(x,t)}).
\end{align}
We find the formal solution of Eq.~\eqref{eq410} in the form
\begin{align}\label{eq420}
Q\ket{\Psi(x,t)}=\int_0^t d\tau e^{i QHQ (t-\tau)/\hbar} QHP\qty(P\ket{\Psi(x,\tau)}),
\end{align}
where we assumed $Q\ket{\Psi(x,0)}=0$ for simplicity.
Inserting Eq.~\eqref{eq420} into Eq.~\eqref{eq400}, we arrive at
\begin{align}\label{eq430}
i\hbar  \pdv{t} \qty(P\ket{\Psi(x,t)}) 
=PHP\qty(P\ket{\Psi(x,t)})+
\int_0^t d\tau PHQ\, e^{i QHQ (t-\tau)/\hbar} \, QHP\qty(P\ket{\Psi(x,\tau)}).
\end{align}
Indeed, the first term on the right-hand side gives a Markovian time evolution due to the system Hamiltonian $PHP$, while the second term gives a non-Markovian time evolution due to the environmental Hamiltonian $QHQ$; see Fig.~\ref{fig4} for a schematic view.
The time derivative of the system state $P\ket{\Psi(x,t)}$ appears to depend on the system state in the past $P\ket{\Psi(x,\tau)}$ directly.
However, the behind-the-scenes story is that the state $P\ket{\Psi(x,\tau)}$ is transferred to the environment by the coupling Hamiltonian $QHP$, evolves in time according to the environmental Hamiltonian $QHQ$ over the period of time $(t-\tau)$, and then is transferred back to the system by the coupling Hamiltonian $PHQ$ at time $t$.
We emphasize that the $E$ dependence of the effective Hamiltonian $\Heff$ in Eq.~\eqref{eq390} corresponds to the non-Markovian memory effect in Eq.~\eqref{eq430}.

\subsection{Non-Hermiticity of the effective Hamiltonian $\Heff$}
\label{subsec3.2}

We now argue that the effective Hamiltonian $\Heff$ in Eq.~\eqref{eq390} appears to be Hermitian, but it is \textit{not} if the environmental Hamiltonian $QHQ$ has an infinite number of degrees of freedom, and as a consequence has a continuous spectrum.
The non-Hermiticity emerges when the value of $E$ falls on the continuous spectrum of $QHQ$, for which the Green's function $(E-QHQ)^{-1}$ is singular.

In such cases, what we do is to introduce an infinitesimal to the denominator of the Green's function.
Specifically, we have retarded and advanced Green's functions in the forms
\begin{align}\label{eq440}
G^\textrm{ret}(E)&:=\frac{1}{E-(QHQ-i\eta)},\\
\label{eq450}
G^\textrm{adv}(E)&:=\frac{1}{E-(QHQ+i\eta)},
\end{align}
where $\eta$ is a positive infinitesimal constant.
We will now give a simple example to show~\cite{Datta95} that the retarded Green's function describes the outgoing waves, Eq.~\eqref{eq30} with $\Re k>0$, while the advanced Green's function describes the incoming waves, Eq.~\eqref{eq30} with $\Re k<0$.

Let us consider the flat space for the environment, for example:
\begin{align}\label{eq455}
QHQ=-\frac{\hbar^2}{2m}\dv[2]{x}
\end{align}
for $\abs{x}>\ell$.
For the retarded Green's function~\eqref{eq440}, the singularity now falls on the  point where the eigenvalue of $QHQ$ is equal to $E+i\eta$.
Assuming that the eigenvalue of the operator~\eqref{eq455} is of the form $\hbar^2{K}^2/{2m}$ for the eigenstate $e^{iKx}$, we have 
\begin{align}\label{eq457}
{K}^2=k^2+i\epsilon,
\end{align}
where $k^2=2mE/\hbar^2$ and $\epsilon=2m\eta/\hbar^2$, which is another positive infinitesimal.
Equation~\eqref{eq457} is further transformed to
\begin{align}
K=k+i\delta\sgn k,
\end{align}
where $\delta=\epsilon/\abs{k}$ is yet another positive infinitesimal.
The eigenstate of $QHQ$  is now given by the form
\begin{align}
e^{iKx}=\exp(ikx-x \delta  \sgn k).
\end{align}
For $x>0$, the solution with $k>0$ is allowed for convergence, which gives a right-going wave,
while for $x<0$ the one with $k<0$ is allowed, which gives a left-going wave.
In short, the pole of the retarded Green's function~\eqref{eq440} is an outgoing wave.
For the advanced Green's function~\eqref{eq450}, on the other hand, we flip the sign of all infinitesimals, and find that its pole is an incoming wave.

This picture is endorsed by the Fourier transform of the Green's functions.
We find
\begin{align}
G^\textrm{ret}(t)&:=\frac{1}{2\pi}\int_{-\infty}^\infty \frac{e^{-iEt/\hbar}}{E-(QHQ-i\eta)} dE
=\begin{cases}
0 & \quad\mbox{for $t<0$},\\
e^{-iQHQt/\hbar} & \quad\mbox{for $t>0$},
\end{cases}
\\
G^\textrm{adv}(t)&:=\frac{1}{2\pi}\int_{-\infty}^\infty \frac{e^{-iEt/\hbar}}{E-(QHQ+i\eta)} dE
=\begin{cases}
e^{-iQHQt/\hbar} & \quad\mbox{for $t<0$},\\
0 & \quad\mbox{for $t>0$},
\end{cases}
\end{align}
which reveals that indeed the former describes the wave propagation out of a source for $t>0$, while the latter describes it into a sink for $t<0$.
This also implies that the former corresponds to the case of the outgoing waves $k>0$ in Eq.~\eqref{eq30} due to the source $\delta(t)$, while the latter to the case of incoming waves $k<0$ due to the sink $\delta(t)$.

We stress here that by taking the retarded and advanced Green's functions in Eq.~\eqref{eq390}, we find the effective Hamiltonians
\begin{align}\label{eq480}
\Heff^\textrm{ret}(E)&:=PHP+PHQ\frac{1}{E-(QHQ-i\eta)}QHP,
\\\label{eq490}
\Heff^\textrm{adv}(E)&:=PHP+PHQ\frac{1}{E-(QHQ+i\eta)}QHP,
\end{align}
explicitly non-Hermitian.
Solving the nonlinear eigenvalue problem~\eqref{eq380} with the retarded effective Hamiltonian~\eqref{eq480} produces the resonant eigenstates, which is the solutions under the outgoing-wave boundary condition, while solving the one with the advanced effective Hamiltonian~\eqref{eq490} produces the anti-resonant eigenstates, which is the solutions under the incoming-wave boundary conditions, both with complex eigenvalues.

In the previous section~\ref{sec2}, the non-Hermiticity is originated in the Siegert boundary condition,
Since the environmental degrees of freedom is eliminated in the Feshbach projection formalism, the non-Hermiticity hidden in the boundary condition came into view in the forms~\eqref{eq480}--\eqref{eq490}.

\subsection{What is the bra vector of the resonant state?}
\label{subsec3.3}

A critical remark is in order.
We observed above that the system Hamiltonian is generally non-Hermitian in the functional spaces of the resonant and anti-resonant states.
In general, a non-Hermitian operator has right- and left-eigenstates, which are generally not Hermitian conjugate to each other as in the case of the Hermitian operators.
Let $J$ denote a non-Hermitian operator; in other words, $J^\dag\neq J$.
The right- and left-eigenstates are given by
\begin{align}
J\ket{\psi_n}=E_n\ket{\psi_n},\\
J^\dag\bra{\phi_n}^\dag={E_n}^\ast\bra{\phi_n}^\dag,
\end{align}
respectively, for which $\ket{\psi_n}\neq\bra{\phi_n}^\dag$ in general.
The right- and left-eigenstates form a complete bi-orthonormal set to span the space as in
\begin{align}
\braket{\phi_m}{\psi_n}=\delta_{mn},
\qquad
\sum_n\dyad{\psi_n}{\phi_n}=I
\end{align}
except at the exceptional points, at which two or more eigenvectors become parallel to each other.

Then one might presume that the probability of a wave vector $\ket{f}$ would be defined as follows:
(i) one would first expand the vector in terms of the right-eigenstates as in
\begin{align}\label{eq530}
\ket{f}=\sum_n f_n\ket{\psi_n};
\end{align}
(ii) one would then construct the corresponding wave vector $\bra{g}$ as in
\begin{align}
\bra{g}=\sum_n {f_n}^\ast \bra{\phi_n};
\end{align}
(iii) then the probability for the system to be in the state $\ket{f}$ would be given by
\begin{align}\label{eq550}
\braket{g}{f}=\sum_n \abs{f_n}^2.
\end{align}
We note that $\bra{g}^\dag\neq\ket{f}$ in general.
The quantity $\qty(\ket{f}^\dag)\ket{f}$ on the contrary would not be a proper definition of the probability.

Nonetheless, we used in Subsection~\ref{subsec2.5} the standard definition of the probability, that is, the square modulus of the wave amplitude, \textit{not} the one defined in the above procedure.
This is actually because the open system with the effective non-Hermitian Hamiltonian is assumed to be a part of a Hermitian closed system~\cite{Kawabata-private}.
Since the probability in the closed system should be given by the square modulus of the wave amplitude, it should be so in a part of it too.
If the system were assumed to be a non-Hermitian closed system, we would define the probability based on the biorthogonal basis~\cite{Brody02} following the procedure given in Eqs.~\eqref{eq530}--\eqref{eq550}. 

\section{Summary}
\label{sec5}

We have reviewed non-Hermitian quantum mechanics of open quantum systems.
The non-Hermiticity emerges in two ways in such systems.
It is implicit if one uses the Siegert boundary condition, but it becomes explicit when we eliminate the environmental degrees of freedom in the Feshbach projection formalism.
In reviewing it, we presented answers to several typical questions that one might ask for non-Hermitian quantum mechanics.
We noted that although the whole set of the eigenstates is time-reversal symmetric, each eigenstate can break the symmetry because the time-reversal operator is not a linear one.
We emphasized that the seemingly Hermitian Hamiltonian can be non-Hermitian depending on the functional space on which the Hamiltonian is defined.
The Hamiltonian is indeed a non-Hermitian operator in the functional space of resonant and anti-resonant states, and hence there is no contradiction in the fact that these states have complex eigenvalues.
We presented a probabilistic interpretation for the resonant and anti-resonant states.
This is thanks to the often misestimated fact that the wave functions of the states spatially diverge.
The spatial divergence physically represents the fact that microscopic objects are attached to the ends of the free space in reality.
We finally note that the use of the standard bra vector (the Hermitian conjugate of the ket vector) is legitimate for non-Hermitian Hamiltonians if they describe open systems that are parts of closed Hermitian systems.

\section*{Acknowledgements}
The study is supported by the Japan Society for Promotion of Science (JSPS) Kakenhi grant Nos.~19H00658, 19F19321 and 21H01005.

\section*{References}
\bibliography{hatano}

\end{document}